**Regional Chaos Synchronization towards controlling high-dimensional open environmental systems**

Yohei Sawada[1]

[1] Department of Civil Engineering, Graduate School of Engineering, the University of Tokyo, Tokyo, Japan

Corresponding author: Y. Sawada, Department of Civil Engineering, the University of Tokyo, Tokyo, Japan, 7-3-1, Hongo, Bunkyo-ku, Tokyo, Japan, yoheisawada@g.ecc.u-tokyo.ac.jp

**Abstract**

Although controlling chaos can be regarded as a form of chaos synchronization, conventional chaos synchronization does not provide a practical framework for controlling high-dimensional open chaotic systems, such as weather. In high-dimensional open environmental systems, it is simply impossible to make the difference between drive and response systems asymptotically approach zero over the entire domain. Instead of conventional *global* chaos synchronization, here I define *regional* chaos synchronization, in which state variables within only a small target region are synchronized with a target trajectory using actuators located in the target region. Theoretical considerations and numerical experiments show that the performance of *regional* chaos synchronization is substantially degraded by the injection of errors from exterior regions that the controller is not intended to modify. Even under the same actuator specifications, the performances of *regional* chaos synchronization differ among trajectories, which is not the case of conventional *global* chaos synchronization. An appropriate target trajectory differs substantially from the natural trajectory only in the target region. This choice of target trajectories can minimize the injection of errors from exterior regions and thereby enable successful *regional* chaos synchronization. The implications of this new framework for realistic control problems such as weather control are discussed.

## 1. Introduction

Chaos synchronization enables multiple chaotic systems that start from different initial conditions to become coupled through a common signal transmitted from one system to another, despite their inherent unpredictability. Since its initial discovery [1], chaos synchronization has provided a foundation of the controllability of chaotic dynamical systems [e.g., 2-3].

This paper focuses on the synchronization of high-dimensional open chaotic systems such as weather. In weather prediction, the connection between data assimilation and chaos synchronization has been discussed [4-6]. Data assimilation integrates observation into numerical simulation to improve estimates of the system state. Although data assimilation methods used operationally in numerical weather prediction systems [7-8] differ substantially from the conventional synchronization methods described in the chaos synchronization literature [1,4,9], data assimilation can be classified as a form of generalized one-way coupled impulsive chaos synchronization [6,10]. Using observations of the global atmosphere as signals, data assimilation can synchronize the state variables of a numerical atmospheric model with the true atmospheric state.

Weather modification or control has long been a grand challenge in meteorology. Since the initial attempt to mitigate a tropical cyclone in the 1960s [11], many ideas have been proposed for mitigating extreme weather events (see Miller et al. 2023 [12] for a comprehensive review). Since the early 2000s, meteorologists have increasingly focused on the strong sensitivity of atmospheric dynamics to initial conditions to identify efficient control methods [13-16]. Miyoshi (2026) [17] explicitly formulated the relationship between chaos synchronization and weather control. In this framework, the trajectory of the natural atmosphere is synchronized with a target, or desired, trajectory by external artificial forcing. Because both data assimilation and weather control can be regarded as forms of chaos synchronization, Miyoshi (2026) [17] discussed their duality. This duality supports recently developed control algorithms which leverage the existing data assimilation methods [18-19].

However, conventional chaos synchronization does not provide a practical framework for controlling weather or many other high-dimensional open systems in the environmental sciences. The formal definition of chaos synchronization can be expressed as:

$$\frac{d\boldsymbol{u}}{dt} = f(\boldsymbol{u}), \qquad \frac{d\boldsymbol{v}}{dt} = f(\boldsymbol{v}) + K(\boldsymbol{u}, \boldsymbol{v}) \tag{1}$$

$$\|\boldsymbol{e}(t)\| = \|\boldsymbol{u}(t) - \boldsymbol{v}(t)\| \to 0, \qquad when\, t \to \infty$$

where $u$ and $v$ are the drive and response systems, respectively. $K$ represents the algorithm of synchronization, and $t$ is time. In weather control, $u$ is the state of desired Earth, e.g, an Earth with reduced precipitation in a specific city, whereas $v$ is that of the actual Earth with human interventions. It is apparently infeasible to design a realistic actuator (i.e., $K$) that achieves (1) in the high-dimensional Earth system. To synchronize a simulated global atmosphere with the actual global atmosphere, data assimilation must substantially modify the simulated state variables (e.g., [20]). Such large-scale modifications can be made in a simulated Earth system but are impossible in the actual Earth system. The duality between data assimilation and control discussed by Miyoshi (2026) [17] therefore implies the impossibility of weather control in the sense of equation (1). Moreover, achieving equation (1) is not the objective of meteorological weather control studies. Rather, meteorologists seek to mitigate specific local phenomena, such as tropical cyclones [16], mesoscale convective systems [21], and isolated convective rainfall events [22], over finite time periods. This objective differs from that expressed in equation (1). Although Miyoshi (2026) [17] also discussed the distinctive characteristics of weather control problems compared with data assimilation, no mathematically rigorous discussion has yet been presented regarding the relationship between chaos synchronization and the practical control of weather and other high-dimensional open environmental systems.

Instead of the conventional *global* chaos synchronization defined in (1), this paper introduces *regional* chaos synchronization as a practical framework for controlling high-dimensional open environmental systems. In *regional* chaos synchronization, the high-dimensional state vectors are decomposed into a targeted region, where the phenomenon of interest occurs, and an exterior region. Only the state vector in the target region is required to synchronize over a finite period:

$$\frac{d\boldsymbol{u}}{dt} = f(\boldsymbol{u}), \qquad \frac{d\boldsymbol{v}}{dt} = f(\boldsymbol{v}) + K(\boldsymbol{u}, \boldsymbol{v}) \tag{2}$$

$$\boldsymbol{u} = \begin{pmatrix} \boldsymbol{u}_\Omega \\ \boldsymbol{u}_{\Omega_c} \end{pmatrix}, \quad \boldsymbol{v} = \begin{pmatrix} \boldsymbol{v}_\Omega \\ \boldsymbol{v}_{\Omega_c} \end{pmatrix}$$

$$\|\boldsymbol{e}_\Omega(t)\| = \|\boldsymbol{u}_\Omega(t) - \boldsymbol{v}_\Omega(t)\| \leq \|\boldsymbol{e}_\Omega(0)\|, \qquad 0 \leq t \leq T$$

where subscripts $\Omega$ and $\Omega_c$ are the target region and its complement exterior regions, respectively. In *regional* chaos synchronization, only the state variables in the target region, $v_\Omega$, are required to synchronize with the drive system. Furthermore, the objective is not to make the error asymptotically approach zero. Instead, *regional* chaos synchronization aims for a finite-time non-amplification of errors in the target region.

Although many recent studies on chaos synchronization have investigated efficient synchronization methods, few have examined the problem formulated in equation (2). Azouani et al. (2014) [9] proposed an efficient and mathematically rigorous algorithm to synchronize two-dimensional Navier-Stokes flow. Wang and Zaki (2022) [23] found that turbulence in channel flow can be synchronized by regional flow velocity observations (and no pressure observation). Fang and Pakzad (2026) [24] showed that two-dimensional Navier-Stokes flow can be synchronized without observing the state near the boundary. Inubushi et al. (2023) [25] identified a sufficiently large observational scale for constraining smaller-scale turbulence in three-dimensional Navier-Stokes flow. These studies explored the use of regional observations, or actuators, to achieve *global* synchronization. In contrast, this study investigates *regional* synchronization using regional observations, or actuators, as a practical control problem in environmental sciences.

## 2. Regional chaos synchronization

In this paper, a simple linear synchronization method is adapted:

$$\frac{d\boldsymbol{u}}{dt} = f(\boldsymbol{u}), \qquad \frac{d\boldsymbol{v}}{dt} = f(\boldsymbol{v}) + \boldsymbol{K}(\boldsymbol{u} - \boldsymbol{v}) \tag{3}$$

where $\boldsymbol{K}$ is the coupling matrix. The synchronization error, $\boldsymbol{e}$, obeys:

$$\frac{d\boldsymbol{e}}{dt} = \left[\boldsymbol{D}f\big(\boldsymbol{u}(t)\big) - \boldsymbol{K}\right]\boldsymbol{e} \tag{4}$$

where $\boldsymbol{D}f\big(\boldsymbol{u}(t)\big)$ is the Jacobian matrix of the chaotic system evaluated along the drive trajectory. For *global* chaos synchronization, all conditional Lyapunov exponents, defined as the Lyapunov exponents of the matrix $\boldsymbol{D}f\big(\boldsymbol{u}(t)\big) - \boldsymbol{K}$, must be negative. As discussed in Section 1, experience of data assimilation in atmospheric science suggests that $\boldsymbol{K}$ cannot be sparse if all conditional Lyapunov exponents are to be made negative. It is therefore infeasible to implement a coupling matrix $\boldsymbol{K}$ that synchronizes the trajectory of the real global atmosphere with an any arbitrary target trajectory in the sense of equation (1).

On the other hand, the linearized dynamics of the synchronization error in the *regional* chaos synchronization setting defined in (2) can be written as:

$$\begin{bmatrix} \frac{de_{\Omega}}{dt} \\ \frac{de_{\Omega_c}}{dt} \end{bmatrix} = \begin{bmatrix} A_{\Omega}(t) - K_{\Omega}(t) & B(t) \\ C(t) & A_{\Omega_c}(t) \end{bmatrix} \begin{bmatrix} e_{\Omega} \\ e_{\Omega_c} \end{bmatrix} \tag{5}$$

$$A_{\Omega}(t) = \frac{\partial \dot{\boldsymbol{u}}_{\Omega}}{\partial \boldsymbol{u}_{\Omega}}, B(t) = \frac{\partial \dot{\boldsymbol{u}}_{\Omega}}{\partial \boldsymbol{u}_{\Omega_c}}, C(t) = \frac{\partial \dot{\boldsymbol{u}}_{\Omega_c}}{\partial \boldsymbol{u}_{\Omega}}, A_{\Omega_c}(t) = \frac{\partial \dot{\boldsymbol{u}}_{\Omega_c}}{\partial \boldsymbol{u}_{\Omega_c}}$$

It should be noted that the coupling, $K_{\Omega}(t)$, is assumed to act only within the targeted region $\Omega$. In other words, the targeted region is always designed to include the locations of the actuators. This is a reasonable assumption because, in spatially continuous systems, actuation must generally be applied near the target phenomenon.

*Regional* chaos synchronization focuses only on the dynamics of synchronization errors in $\Omega$:

$$\frac{de_{\Omega}}{dt} = \big(A_{\Omega}(t) - K_{\Omega}(t)\big)e_{\Omega} + B(t)e_{\Omega_c} \tag{6}$$

Let $\Phi_{\Omega}(t,s)$ be the state transition matrix of the dynamics of $\dot{z} = \big(A_{\Omega}(t) - K_{\Omega}(t)\big)z$. Then, the equation (6) gives:

$$e_{\Omega}(t) = \Phi_{\Omega}(t,0)e_{\Omega}(0) + \int_0^t \Phi_{\Omega}(t,s)\, B(s)e_{\Omega_c}(s)ds \tag{7}$$

The first term on the right-hand side of (7) represents the growth or suppression of the initial error in the target region, while the second term represents the propagation of errors from the external region, where no control intervention is applied. I assume that:

$$\|\Phi_{\Omega}(t,s)\| \le e^{-\gamma(t-s)}, \|B(t)\| \le b, \left\|e_{\Omega_c}(t)\right\| \le \left\|e_{\Omega_c}(0)\right\|e^{\lambda t} \tag{8}$$

where $\gamma > 0$ is the regional suppression rate, and $\lambda > 0$ is the error growth rate of the original (driver) dynamics. Consequently,

$$\|e_{\Omega}(t)\| \le e^{-\gamma t}\|e_{\Omega}(0)\| + \frac{b}{\gamma + \lambda}\big(e^{\lambda t} - e^{-\gamma t}\big)\left\|e_{\Omega_c}(0)\right\| \tag{9}$$

Equation (9) indicates that the synchronized error in $\Omega$ does not asymptotically approach zero, even when the regional suppression rate is sufficiently large, because errors in the target region are amplified by the influx of errors from the uncontrolled exterior region, where the initial error exponentially grows.

In *regional* chaos synchronization, the goal is the finite-time non-amplification of the regional synchronization error. How long can the error be maintained below its initial magnitude? In this paper, control window, $T_{control}$, is used as an indicator of successful control. From (9) and assuming a large regional suppression rate such that $e^{-\gamma T_{control}} \approx 0$, $T_{control}$ can be approximated as:

$$T_{control} \approx \frac{1}{\lambda} \log\left(\frac{(\gamma + \lambda)\|e_{\Omega}(0)\|}{b\|e_{\Omega_c}(0)\|}\right) \tag{10}$$

Equation (10) highlights several key factors for successful control. First, a slower exterior error growth rate, $\lambda$, extends the control window, although this parameter cannot be altered because it is determined by the error growth rate of the original uncontrolled dynamics. Second, a larger regional suppression rate, $\gamma$, extends the control window, which is straightforward. Third, weaker exterior-to-interior coupling, represented by $b$, extends the control window. Because this coupling depends on local dynamics, the target region and target trajectory can be chosen to minimize $b$, e.g., targeting isolated convective cells which are not strongly influenced by external large-scale flows. Finally, the control window is extended when the initial error in the target region is larger than that in the external region (i.e., $\|e_{\Omega}(0)\| > \|e_{\Omega_c}(0)\|$). It is therefore advantageous to choose a target trajectory whose significant difference to the natural trajectory appears only in the target region.

## 3. Numerical experiment

### 3.1. Experiment design

To demonstrate *regional* chaos synchronization, a numerical experiment was conducted using the two-dimensional Kuramoto-Sivashinsky-type equation:

$$\frac{du}{dt} = -\frac{1}{2}\left(\left(\frac{du}{dx}\right)^2 + \left(\frac{du}{dy}\right)^2\right) - \Delta u - \Delta^2 u \tag{11}$$

where $u$ is the state. Equation (11) was solved using a Fourier pseudo-spectral method with two-thirds dealiasing. Time integration was performed with a fourth-order exponential time-differencing Runge-Kutta scheme with the time stepping of 0.01. The square periodic domain $(x, y)$ consisted of $256 \times 256$ with a spatial resolution of 0.5. The initial condition was specified as:

$$u(x, y) = 0.1\cos(4\pi x/L) + 0.1\sin(6\pi y/L) + \varepsilon \tag{12}$$
$$\varepsilon \sim N(0,1)$$

where L is the domain size (=128).

To prepare the chaos synchronization experiments, the equation (11) is integrated until $t = 70$. Gaussian while noise drawn independently from $N(0,0.1)$ was then added to the state. The original trajectory was treated as the drive system ($u$), whereas the trajectory initiated from the perturbed state was treated as the response system ($v$). This procedure was repeated 100 times by independently drawing $\varepsilon$ in equation (12) (i.e., initial conditions of the driver system at $t = 0$) thereby generating 100 pairs of drive and response trajectories.

Both the drive and response systems were integrated without control forcing from $t = 70$ to $t = 80$ to eliminate unphysical errors induced by the while noise. Figure 1 shows an example of the states of the drive and response systems at the start of chaos synchronization (i.e., $t = 80$). In this paper, a simple nudging method was adopted:

$$\frac{d\boldsymbol{v_i}}{dt} = f(\boldsymbol{v}) + K_i(u_{\boldsymbol{i}} - v_{\boldsymbol{i}}) \tag{13}$$

where $v_i$ is the *i*-th element of the response state, $\boldsymbol{v}$. $f$ represents the two-dimensional Kuramoto-Sivashinsky equation given by (11). At the actuator locations, $K_i > 0$; otherwise, $K_i = 0$. There are two hyperparameters in the synchronization experiment: actuator density and gain. The actuators were placed at intervals of 2, 4, or 6 grid points in both the x and y directions. The gains, $K_i$, were set to 0.03, 0.06, or 0.09. Two synchronization scenarios were considered: global and regional synchronization. The target region, $\Omega$, was defined as the rectangular region shown in Figure 1 ($60 \leq x \leq 68, 60 \leq y \leq 68$). In the *global* synchronization experiments, actuators are placed throughout the entire domain. In contrast, in the *regional* synchronization experiments, the actuators were placed only in $\Omega$. Three actuator-density settings, three gain settings, and two synchronization scenarios were considered, resulting in a total of 18 experiments for each initial condition. These 18 experiments were repeated for each of the 100 pairs of drive and response systems.

We evaluated the error, $\|e_\Omega(t)\|$, as the root-mean-square error (RMSE) relative to the drive system in $\Omega$. Only grid points without actuators were used to calculate the RMSE. To explain the performance of *regional* chaos synchronization, the initial error ratio of the interior regions to the exterior regions $\left(\frac{\|e_\Omega(0)\|}{\|e_{\Omega_c}(0)\|}\right)$ is computed. For this calculation, the exterior region, $\Omega_c$, was defined as the region surrounding $\Omega$ ($50 \leq x \leq 78, 50 \leq y \leq 78$), because the interactions between interior and exterior regions occur across their boundary. The regional suppression rate, $\gamma$, was estimated from the timeseries of $\|e_\Omega(t)\|$ in the *global* chaos synchronization experiments using regression over the period from $t = 80$ to $t = 85$. It was assumed that the same actuator configuration, in terms of density and gain, produced the same regional suppression rate for both global and regional synchronization experiments. Finally, the control window, $T_{control}$, was computed as the first positive time when $\|e_{\Omega}(t)\|$ exceeded $\|e_{\Omega}(0)\|$. If $\|e_{\Omega}(t)\|$ never decreased below $\|e_{\Omega}(0)\|$, $T_{control}$ was set to 0.

**3.2. Results**

Figure 2 illustrates the synchronized experiments. When the actuator density and gain are sufficiently large, *global* synchronization can be achieved. However, even at the same actuator density and gain, when actuators are placed only in $\Omega$, the error can be suppressed only for a short period. A regional error suppression rate comparable to that achieved in global synchronization is observed only during the first few time steps, after which the error grows rapidly because of error injection from $\Omega_c$ (see equation (7)). This result clearly indicates that finite-time non-amplification of the error is a reasonable goal for *regional* chaos synchronization. Figure 2d shows the relationship between regional suppression rate ($\gamma$) and control window ($T_{control}$). The simple logarithmic relationship implied by equation (10) approximately holds.

However, this clear relationship between the regional suppression rate and the control window is not observed across the 100 pairs of drive and response systems. Figure 3a reproduces the relationship shown in Figure 2d for all 100 experiments, each with a different initial condition for the drive system. Figure 3a indicates that the regional suppression rate does not substantially vary among the 100 experiments under the same actuator configuration. For instance, when the interval and gain of $K_i$ are set 2 and 0.09, respectively, the regional suppression rate ranges from 1.5 to 2.5 in most cases (see the red triangles in Figure 3a). This result implies that the performance of *global* chaos synchronization does not greatly vary among the 100 drive-response pairs under this configuration. In contrast, the performance of *regional* chaos synchronization, quantified by control window, substantially differs among the drive-response pairs. This finding suggests that the success or failure of *regional* chaos synchronization strongly depends on the trajectories of driver (target) and response (real world) systems.

To consider the difference among the trajectories in each experiment, another metric based on equation (10) was developed to characterize the efficiency of control. Figure 10b shows the relationship between $\gamma \frac{\|e_{\Omega}(0)\|}{\|e_{\Omega_c}(0)\|}$ and control window ($T_{control}$) and indicates that incorporating differences between initial errors in the interior and exterior

regions helps explain the variation in performance among the experiments. When $\|e_{\Omega}(0)\| < \|e_{\Omega_c}(0)\|$, errors are advected from the exterior region into the interior, or target, region. The advection of errors that are larger in the exterior region than in the target region strongly hinders the sustained suppression of errors within the target region. This is why a large regional suppression rate does not necessarily ensure successful *regional* chaos synchronization. When $\|e_{\Omega}(0)\| > \|e_{\Omega_c}(0)\|$, errors in the target region are not initially amplified by the advection of error from the exterior regions. Consequently, the error suppression theoretically predicted by $\gamma$ is realized during the initial period of the experiment. Note that even in this case, the non-amplification of errors in the target region eventually fails because of error advection from the uncontrolled exterior regions, where the error rapidly grows. It should also be noted that the metric $\gamma \frac{\|e_{\Omega}(0)\|}{\|e_{\Omega_c}(0)\|}$, inspired by equation (10), cannot perfectly predict $T_{control}$. This is because local and flow-dependent error advection (characterized by $B(t)$ and $C(t)$ in equation (5)) is not considered. Moreover, the linearization of the error dynamics, which is a fundamental assumption underlying the parameterization in equation (10), may not remain fully valid throughout the control window.

## 4. Conclusions and discussion

In this paper, I present *regional* chaos synchronization as a framework of realistic control problems in high-dimensional open systems such as weather. While conventional *global* chaos synchronization aims to make the differences in all state variables between drive and response systems asymptotically approach zero, *regional* chaos synchronization aims to achieve finite-time error non-amplification of errors within a limited target region under the extremely stringent constraints on the number and magnitude of control interventions. Theoretical considerations and numerical experiments show that the performance of *regional* chaos synchronization is substantially degraded by the injection of errors from exterior regions that the controller is not intended to modify. Even with the same actuator configuration, the performance of *regional* chaos synchronization varies among trajectories, unlike that of conventional *global* chaos synchronization. This strong trajectory dependence reveals that the choice of an target trajectory is as important as the design of the actuator to maximize the regional suppression rate. When a target region is defined as a small region where the phenomenon of interest occurs, an appropriate target trajectory should substantially differ from the natural trajectory only in the target region.

Selecting such a target trajectory can minimize the injection of errors from exterior regions and thereby lead successful *regional* chaos synchronization.

The importance of the choice of an appropriate target trajectory was also emphasized by [17], who introduced the concept of modifiability. A target trajectory should be chosen from within a modifiable range, defined as the set of all possible future trajectories that can be reached by available interventions. The present study of *regional* chaos synchronization provides additional insight into the selection of appropriate target trajectories. The target trajectory should have a modifiable difference to nature with respect to the target phenomenon, while avoiding large differences outside the target region.

Another important insight is that finite-time non-amplification of errors is a reasonable goal of weather control. This implies that the target trajectory must be switched adaptively to maintain the system in a desirable state. Recently proposed ensemble-based control methods, although not explicitly designed for this purpose, possess this property. Miyoshi and Sun (2022) [15] identify the "best" and "worst" ensemble members from an ensemble prediction and determine control interventions based on the difference between them. They update the "best" trajectory whenever control interventions are computed. Although ensemble Kalman control [16,18] and its variant [26] do not explicitly define a target trajectory, they implicitly use a target trajectory represented by a linear combination of ensemble members. These existing algorithms adaptively change the target trajectory because the "optimal" trajectories are different in the different timings. Also, due to the limited predictive skill of ensemble, the estimated "optimal" trajectory should be updated as the lead time to the occurrence of the target phenomenon decreases. The existing weather control algorithms suggest switching target trajectories because the user-defined optimality of a trajectory changes in time. In contrast, *regional* chaos synchronization suggests switching target trajectories simply because it is impossible to keep synchronizing with a single pre-defined trajectory. The existing ensemble-based weather control algorithms do not consider modifiability in the sense discussed by [17] or in the sense introduced in the present study, when they adaptively choose target trajectories. Future work should focus on the algorithms to explore appropriate target trajectories in the *regional* chaos synchronization.

**Acknowledgements**

This work was supported by the JST Moonshot R&D program (Grant JMPJMS2281).

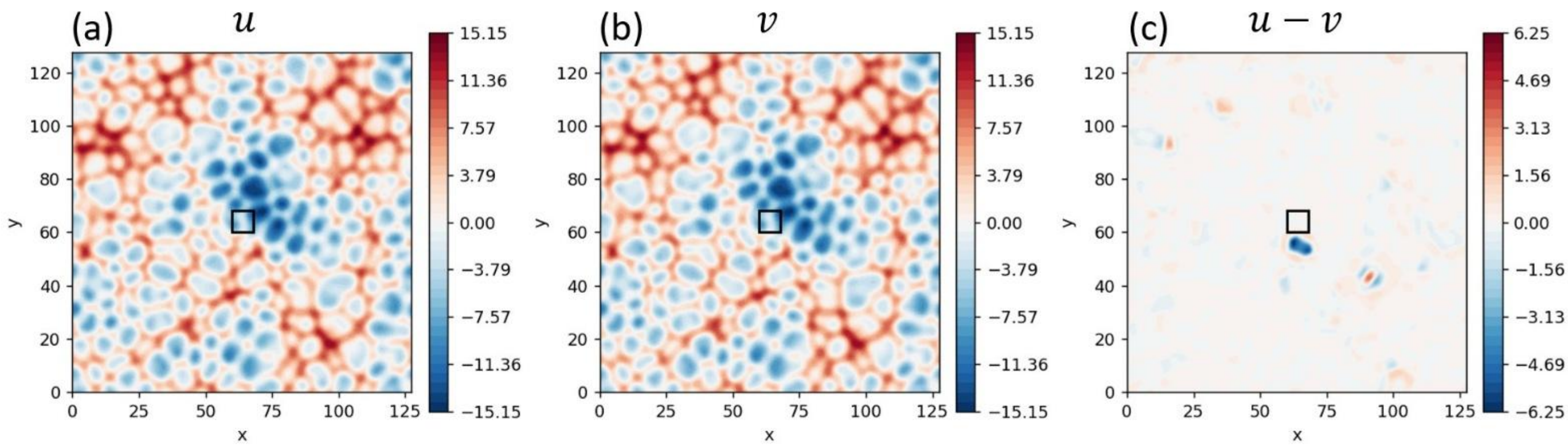

**Figure 1.** An example of the initial conditions of chaos synchronization experiments. The state at $t = 80$ of (a) drive system, (b) response system, and (c) their difference. Black boxes show the target region. See Section 3.1 for details.

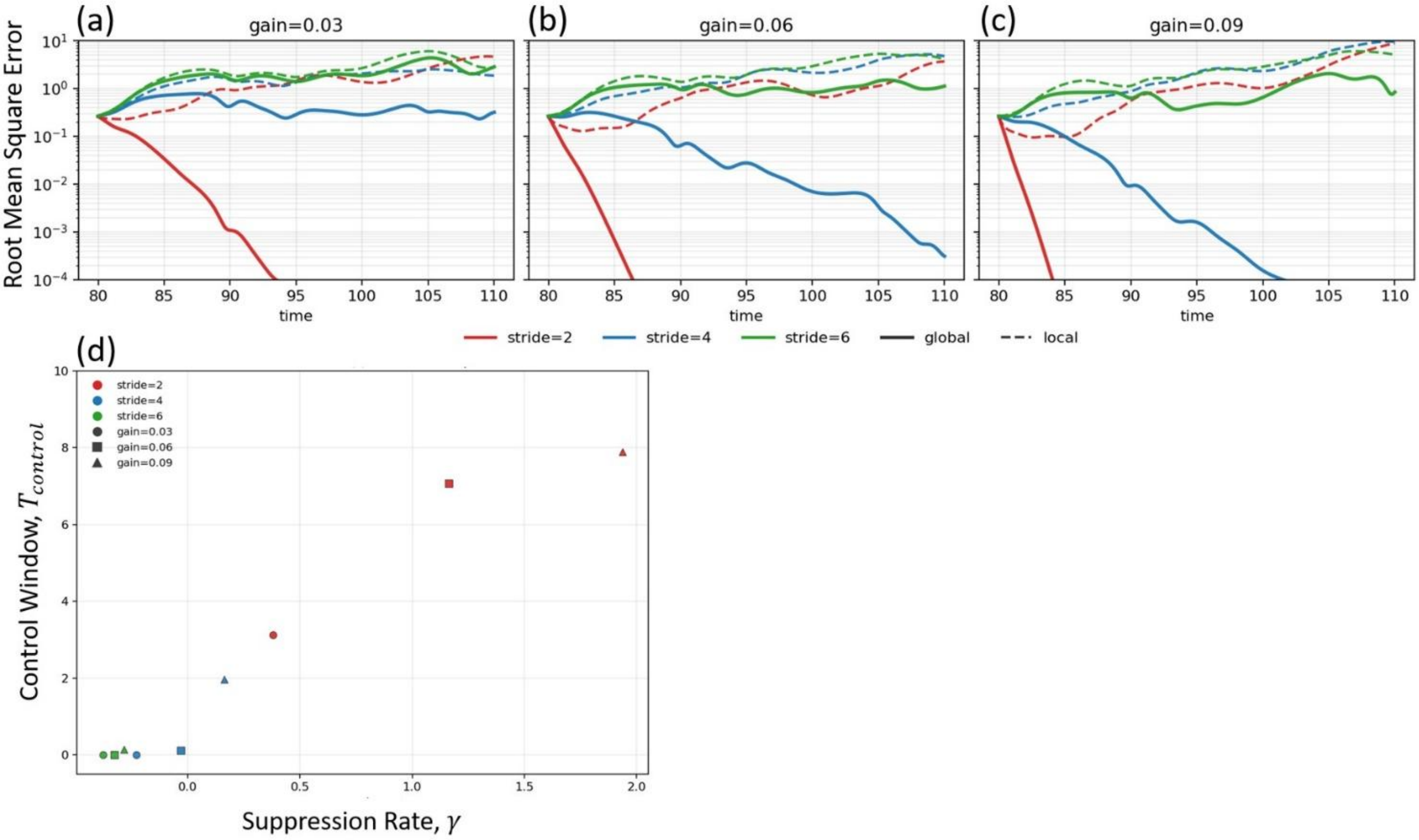

**Figure 2.** An example of synchronization experiments. (a-c) RMSE against the drive system in the target region ($\|e_\Omega(t)\|$) in global (solid lines) and regional (dashed lines) synchronization experiments by actuator intervals of 2 (red), 4 (blue), and 6 (green) and actuator gains of (a) 0.03, (b) 0.06, and (c) 0.09. (d) The relationship between suppression rate and control window in regional synchronization experiments by actuator intervals of 2 (red), 4 (blue), and 6 (green) and actuator gains of 0.03 (circles), 0.06 (squares), and 0.09 (triangles).

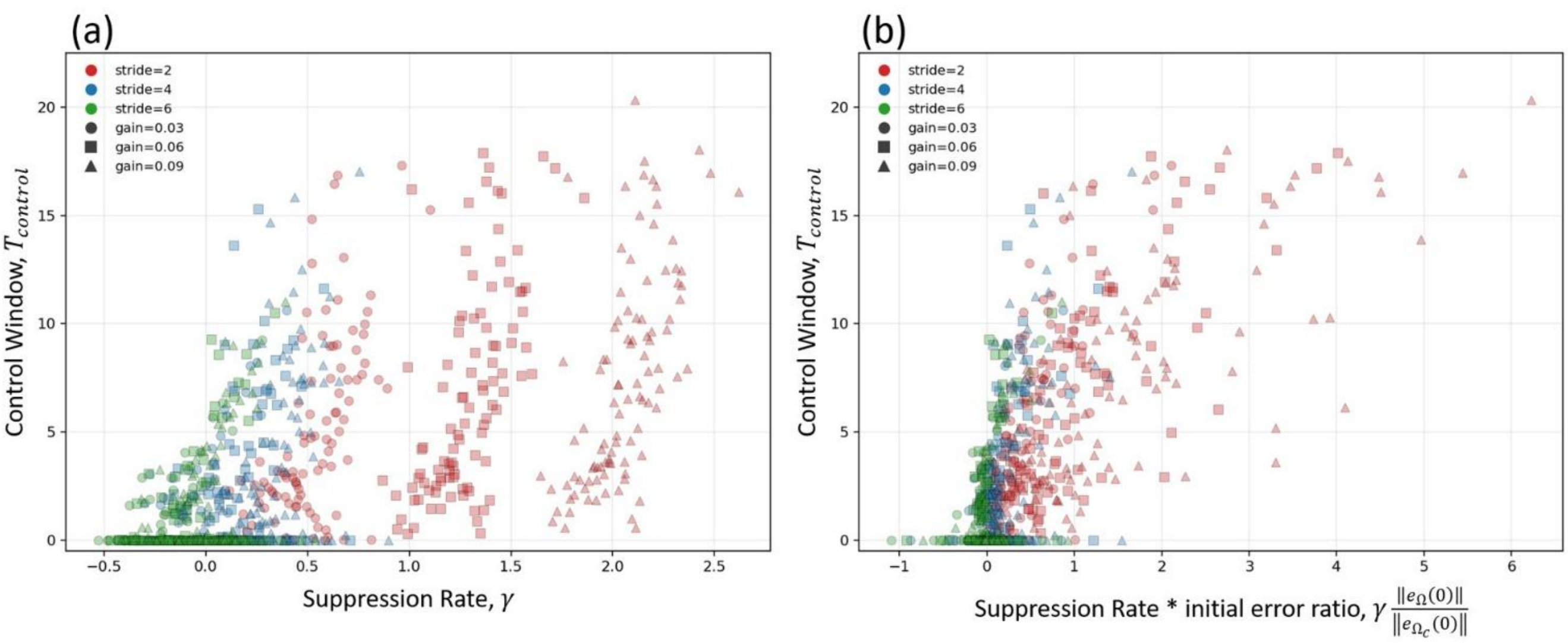

**Figure 3.** (a) same as Figure 2d, but for all 100 pairs of drive and response systems. (b) same as (a) but horizontal axis is replaced by $\gamma \frac{\|e_\Omega(0)\|}{\|e_{\Omega_c}(0)\|}$.